\documentclass[twocolumn]{aastex63}

\usepackage{natbib}

\usepackage{graphicx}
\usepackage{epstopdf}
\usepackage{amsmath}
\usepackage{amssymb}
\usepackage{bm}
\usepackage{color}
\usepackage{multirow}
\usepackage{rotating}
\usepackage{makecell}

\def\gs{\mathrel{\raise0.35ex\hbox{$\scriptstyle >$}\kern-0.6em
\lower0.40ex\hbox{{$\scriptstyle \sim$}}}}
\def\ls{\mathrel{\raise0.35ex\hbox{$\scriptstyle <$}\kern-0.6em
\lower0.40ex\hbox{{$\scriptstyle \sim$}}}}

\newcommand{\mbh}{$M_{\rm BH}$}

\newcommand{\Hb}{H$\beta$}

\newcommand{\civ}{C\,{\sc iv}}

\newcommand{\ciii}{C\,{\sc iii}]}

\defcitealias{Williams++20b}{Paper~I}

\newcommand{\parlogmbh}{8.31^{+0.07}_{-0.06}}
\newcommand{\parfellip}{0.83^{+0.04}_{-0.06}}
\newcommand{\parfflow}{0.26^{+0.18}_{-0.18}}
\newcommand{\parthetae}{20.2^{+13.6}_{-13.2}}

\newcommand{\parrmedian}{33.0^{+2.4}_{-2.1}}
\newcommand{\parrmin}{8.2^{+1.3}_{-1.1}}

\newcommand{\partaumedian}{36.4^{+1.8}_{-1.8}}
\newcommand{\parbeta}{1.32^{+0.11}_{-0.09}}
\newcommand{\parthetao}{52.6^{+5.5}_{-7.8}}
\newcommand{\parthetai}{40.8^{+5.6}_{-6.5}}
\newcommand{\parkappa}{< -0.47}
\newcommand{\pargamma}{2.6^{+1.4}_{-1.2}}
\newcommand{\parxi}{> 0.77}

\newcommand{\fmeanfwhm}{-0.32^{+0.09}_{-0.07}}
\newcommand{\fmeansigma}{0.20^{+0.09}_{-0.07}}
\newcommand{\frmsfwhm}{-0.46^{+0.10}_{-0.09}}
\newcommand{\frmssigma}{-0.08^{+0.09}_{-0.07}}

\shorttitle{Modeling the \civ\ BLR of SDSS J2222+2745} 
\shortauthors{Williams et al.}
 
\begin{document}

\title{Dynamical Modeling of the \civ\ Broad Line Region of the ${\bm z}~\mathbf{=2.805}$ Multiply Imaged Quasar SDSS J2222+2745}

\author[0000-0002-4645-6578]{Peter R. Williams}
\affiliation{Department of Physics and Astronomy, University of California, Los Angeles, CA 90095-1547, USA}

\author[0000-0002-8460-0390]{Tommaso Treu}
\altaffiliation{Packard Fellow}
\affiliation{Department of Physics and Astronomy, University of California, Los Angeles, CA 90095-1547, USA}

\author[0000-0003-2200-5606]{H\r{a}kon Dahle}
\affiliation{Institute of Theoretical Astrophysics, University of Oslo, PO Box 1029, Blindern 0315, Oslo, Norway}

\author[0000-0001-8818-0795]{Stefano Valenti}
\affiliation{Department of Physics, University of California, Davis, CA 95616, USA}

\author[0000-0002-8860-1032]{Louis Abramson}
\affiliation{The Observatories of the Carnegie Institution for Science, 813 Santa Barbara St., Pasadena, CA 91101, USA}

\author[0000-0002-3026-0562]{Aaron J. Barth}
\affiliation{Department of Physics and Astronomy, University of California at Irvine, 4129 Frederick Reines Hall, Irvine, CA 92697-4575, USA}

\author{Brendon J. Brewer}
\affiliation{Department of Statistics, The University of Auckland, Private Bag 92019, Auckland 1142,
New Zealand}

\author{Karianne Dyrland}
\affiliation{Institute of Theoretical Astrophysics, University of Oslo, PO Box 1029, Blindern 0315, Oslo, Norway}
\affiliation{Kongsberg Defence \& Aerospace AS, Instituttveien 10, PO Box 26, 2027, Kjeller, Norway}

\author[0000-0003-1370-5010]{Michael Gladders}
\affiliation{Department of Astronomy \& Astrophysics, The University of Chicago, 5640 S. Ellis Avenue, Chicago, IL 60637, USA}

\author[0000-0003-1728-0304]{Keith Horne}
\affiliation{SUPA Physics and Astronomy, University of St. Andrews, Fife, KY16 9SS, UK}

\author[0000-0002-7559-0864]{Keren Sharon}
\affiliation{Department of Astronomy, University of Michigan, 1085 S. University Avenue, Ann Arbor, MI 48109, USA}

\correspondingauthor{Peter R. Williams}
\email{pwilliams@astro.ucla.edu}

\begin{abstract}
  We present \civ\ BLR modeling results for the multiply imaged $z=2.805$ quasar SDSS J2222+2745.
  Using data covering a 5.3 year baseline after accounting for gravitational time delays, we find models that can reproduce the observed emission-line spectra and integrated \civ\ fluctuations.
  The models suggest a thick disk BLR that is inclined by $\sim$40 degrees to the observer's line of sight and with a emissivity weighted median radius of $r_{\rm median} = \parrmedian$ light days.
  The kinematics are dominated by near-circular Keplerian motion with the remainder inflowing.
  The rest-frame lag one would measure from the models is $\tau_{\rm median} = \partaumedian$ days, which is consistent with measurements based on cross-correlation.
  We show a possible geometry and transfer function based on the model fits and find that the model-produced velocity-resolved lags are consistent with those from cross-correlation. 
  We measure a black hole mass of $\log_{10}(M_{\rm BH}/M_\odot) = \parlogmbh$, which requires a scale factor of $\log_{10}(f_{{\rm mean},\sigma}) = \fmeansigma$.
\end{abstract}



\section{Introduction}
\label{sect:intro}

Precise measurements of supermassive black hole (BH) masses across cosmic time are a necessary ingredient for understanding BH formation and growth and the connection between BHs and their host galaxies \citep{ferrarese00,gebhardt00,Ding++20}.
The most successful technique for measuring \mbh\ outside the local universe is reverberation mapping \citep[RM,][]{blandford82, peterson93, peterson14, ferrarese05}, which measures the response of the broad emission-line region (BLR) to changes in the continuum.
By combining emission-line widths with the time lag between continuum and emission-line fluctuations, one can obtain a virial estimate of the black hole's mass:
\begin{align}
M_{\rm BH} = f\frac{c\tau(\Delta V)^2}{G}.
\end{align}
\label{eq:revmap}
The scale factor, $f$, accounts for the geometry, kinematics, and orientation of the BLR, which are intrinsic to each individual region and, in general, unknown.
This is the largest source of uncertainty in RM \mbh\ measurements, estimated to be $\sim$0.4 dex \citep{park12b}.

Direct modeling of the BLR \citep{pancoast11, pancoast12, brewer11b} avoids this issue entirely by including \mbh\ as a free parameter.
Additionally, models inform us of the structure and kinematics of the BLR and can provide $f$ values for individual BLRs.
This opens the possibility of finding correlations between $f$ and other observables that could be used to improve \mbh\ measurements for all active galactic nuclei (AGNs), not just those with data suitable for modeling \citep[][Villafana et al. 2021, in prep.]{Williams++18}.

Until now, the dynamical modeling approach has been applied only to nearby AGNs with $z < 0.1$, and only one of these analyses \citep[NGC 5548,][]{Williams++20a} examined the UV-emitting BLR.
This is due to the complexities of high-$z$ reverberation mapping campaigns paired with the high signal-to-noise ratio ($S/N$) spectra required for modeling.
However, studies of BH growth require precise \mbh\ measurements across all stages of the Universe, so an understanding of the UV BLR in the early Universe and in quasar-like environments is necessary.

The extraordinary data set of the multiply imaged quasar SDSS J2222+2745 \citep[discovered by][]{Dahle++13} described by \citet[][hereafter Paper~I]{Williams++20b} is the first high-$z$ monitoring campaign with data quality good enough for the modeling approach.
In this paper, we model the \civ-emitting BLR in SDSS J2222+2745 using the data presented in \citetalias{Williams++20b}.
In Section \ref{sect:dataandmethods}, we briefly describe the data set and the BLR modeling approach used in the analysis.
In Section \ref{sect:results}, we present the model fits to the data and describe the inferred BLR geometry and kinematics.
We also compute the scale factor $f$ for the SDSS J2222+2745 \civ\ BLR and compare it to values for the \Hb\ BLRs of other AGNs determined using the same modeling approach. 
We conclude in Section \ref{sect:conclusions}.
When necessary, we adopt a $\Lambda$CDM cosmology with $H_0 = 70~{\rm km~s}^{-1}~{\rm Mpc}^{-1}$, $\Omega_M = 0.3$, and $\Omega_\Lambda = 0.7$.


\section{Data and methods}
\label{sect:dataandmethods}
The data used in this analysis are the same data presented in \citetalias{Williams++20b}, and the modeling approach is described in detail by \citet{pancoast14b}.
Here, we briefly summarize both components, but direct the reader to the respective papers for detailed explanations.

\subsection{Spectroscopic and photometric data}
\label{sect:data}
Beginning June 2016, we obtained monthly spectra of the three brightest images of SDSS J2222+2745 with the Multi-Object Spectrograph at Gemini Observatory North \citep[GMOS-N;][]{GMOS}.
The spectra covered $\sim$5000 to 8200 \AA\ after dithering and flux calibration, covering the \civ\ and \ciii\ broad emission lines.
In addition, we obtained $g$-band photometry with roughly twice-per-month cadence, beginning in September 2011 with the Alhambra Faint Object Spectrograph and Camera (ALFOSC) at the 2.56m Nordic Optical Telescope (NOT).

Since we are modeling only the emission of the \civ\ BLR, we first need to isolate the \civ\ broad emission line from the other spectral components.
As described in \citetalias{Williams++20b}, we model \civ\ as a fourth-order Gauss-Hermite polynomial, and we use those fits in this analysis.
We combine the time-series of spectra for each image by first multiplying the fluxes by the corresponding image magnification and then shifting the times by the measured time delays \citep[$\Delta\tau_{AB} = -42.44$ days and $\Delta\tau_{AC} = 696.65$ days;][]{Dyrland19}, setting image A as the reference.

The 2020 spectra for the leading image C suffer from low $S/N$ as a result of the small image magnification paired with relaxed observing condition constraints in 2020 due to the COVID-19 pandemic.
For this reason, we remove these data from the analysis.

To speed up the modeling code, we also combine spectra that fall within 7 days of each other (observed frame), after accounting for gravitational time delays.
This corresponds to $\sim$2 days in the rest frame, which is over an order of magnitude smaller than the expected BLR size.
Since our smooth BLR model would be unable to resolve variations on this timescale, we lose no constraining power by combining these spectra.
After making these changes, our final data set consists of 63 spectra covering a 1944 day (5.3 year) baseline, after accounting for gravitational time delays.

Finally, we bin the individual spectra by a factor of 8 in wavelength to further decrease computation time, giving 93 wavelength bins from 5503.5 to 6239.5 \AA.
The smooth BLR model is unable to resolve emission-line structure on the 1 \AA/pix (50 km/s/pix) scale of the raw data, and we are using smooth Gauss-Hermite fits to \civ, so attempting to fit all pixels would introduce unnecessary computational burden while providing no additional constraining power.

\subsection{BLR Model}
We model the BLR emission using a collection of point particles surrounding the central black hole and ionizing source.
The ionizing light propagates outwards and as it reaches the particles, they instantaneously re-process the light and emit it towards the observer in the form of emission lines.

The positions and velocities of the particles are determined by a number of free parameters, described in detail by \citet{pancoast14b}.
To summarize, the particles are distributed in a thick disk with half-opening angle $\theta_o$ and inclined relative to the observer's line of sight by $\theta_i$ ($\theta_i = 0$ deg is face-on).
An additional parameter, $\gamma$, determines if the particles are distributed uniformly throughout the thick disk ($\gamma = 1$) or if they are concentrated at the faces of the disk ($\gamma = 5$).
The distances of the particles from the origin are drawn from a Gamma distribution with shape parameter $\beta$, and mean $\mu$, that has been shifted from the origin by a minimum radius $r_{\rm min}$.

A parameter $\kappa$ determines the relative brightness of each particle by controlling if particles re-emit preferentially back towards the origin ($\kappa = -0.5$), isotropically ($\kappa = 0$), or away from the origin ($\kappa = 0.5$).
The disk midplane can range from being fully opaque ($\xi = 0$) to fully transparent ($\xi = 1$).

A fraction of the particles, $f_{\rm ellip}$, are assigned to have near-circular orbits with radial and tangential velocities drawn from a distribution centered on the circular velocity.
The remaining particles all have either inflowing or outflowing trajectories based on the binary parameter $f_{\rm flow}$ ($<0.5$ inflow, $>0.5$ outflow).
The radial and tangential velocities are drawn from a distribution centered on the radial escape velocity that is rotated by an angle $\theta_e$ in the $v_r - v_\phi$ plane. 
This allows for particles with purely radial motion ($\theta_e\sim 0$ deg) or on highly elliptical, bound orbits ($\theta_e\sim 45$ deg).

When interpreting the model parameters, it is important to keep in mind that we are modeling the BLR emission rather than the underlying gas. 
We do not include in our model the complex photoionization process, which would require additional assumptions about the BLR environment---such as the gas density, temperature, and metallicity distributions---and the relation between the observed $g$-band continuum and the ionizing spectrum.

\subsection{Fitting the model to data}
Using the observed continuum light curve, we can compute the time-series of emission-line spectra that a given model produces by summing the contributions of each BLR particle, taking into account the position-induced time lag and velocity-induced wavelength shift.
Our goal is to explore the model parameter space such that the model-produced spectra best fit the observed spectra.
We assume that the ionizing continuum follows the observed $g$-band continuum and use Gaussian processes as a way to flexibly interpolate between observed data points.

We use a Gaussian likelihood function to compare the emission-line spectra with the model spectra.
In post-processing, we also introduce a statistical temperature, $T$, that softens the likelihood function, effectively increasing the spectra uncertainties.
This accounts for under-estimates of uncertainties as well as the challenge of fitting a complex BLR with a simple model.
To explore the parameter space of both the BLR model and continuum model, we use the diffusive nested sampling code {\sc DNest4} \citep{dnest4}.
Diffusive nested sampling is a modification of the nested sampling technique that is particularly efficient at exploring complex, high-dimensional probability spaces.
From the code output, we produce a posterior sample from which we can infer the model parameter values.

\section{Results}
\label{sect:results}

\begin{figure}[h!]
\begin{center}
\includegraphics[width=3.3in]{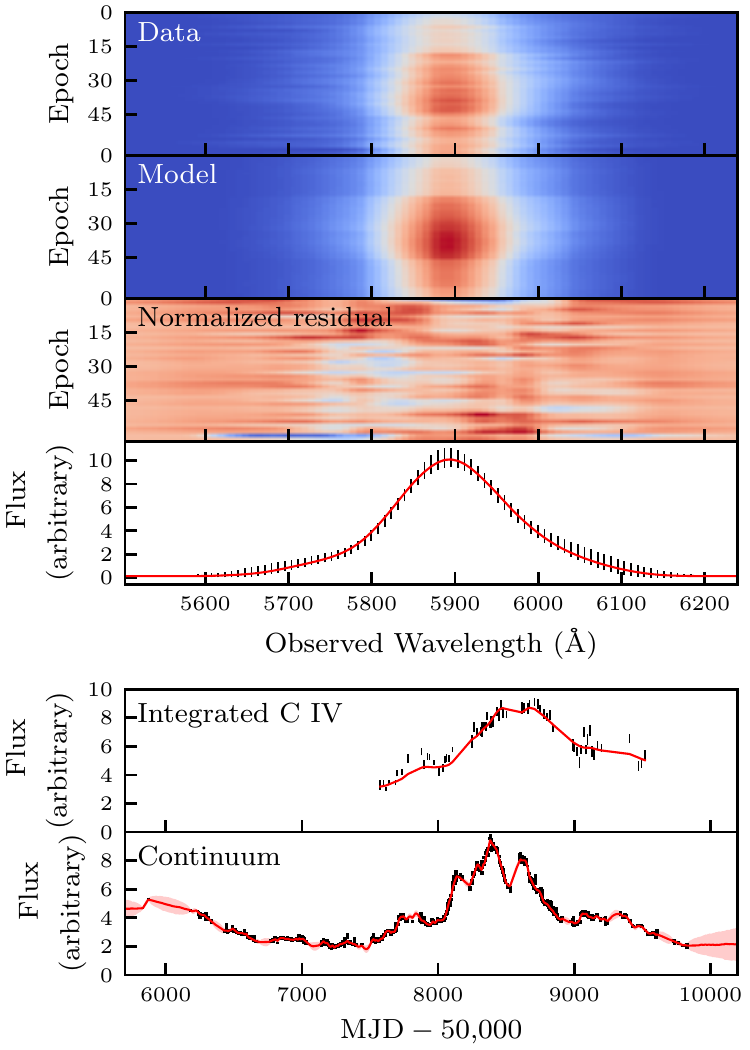}  
\caption{
From top to bottom, 
\textit{Panel 1}: Time-series of observed spectra.
\textit{2}: One possible model fit to the observed spectra.
\textit{3}: Normalized residual ($[{\rm Data - Model}]/{\rm Data~uncertainty}$).
\textit{4}: Observed spectrum from one of the epochs with uncertainties multiplied by $\sqrt{T}$ (black points) and the spectrum produced by the model shown in panel 2 (red).
\textit{5}: Integrated \civ\ emission-line flux of the data (black points) and model (red).
\textit{6}: Observed $g$-band continuum light curve (black) and the model fit to the light curve (red).
\label{fig:display}}
\end{center}
\end{figure}

The modeling code is able to find regions of the BLR model parameter space that reproduce the observed emission-line shape and fluctuations (Figure \ref{fig:display}).
To avoid over-fitting, we soften the likelihood function with a temperature $T = 180$ when constructing the posterior sample from the {\sc DNest4} output.
This is equivalent to multiplying all uncertainties on the spectra by a factor of $\sqrt{T} = 13.4$, which is reflected in the error bars in Panels 4 and 5.
The model displayed in Figure \ref{fig:display} was chosen to be representative of the full posterior sample and has parameter values close to those reported in Section \ref{sect:params}.
The large-scale variations in the \civ\ emission-line flux are very well captured by this model, although some of the shorter-timescale fluctuations ($<$100 days) are smoothed out.

\begin{figure}[h!]
\begin{center}
\includegraphics[width=3.2in]{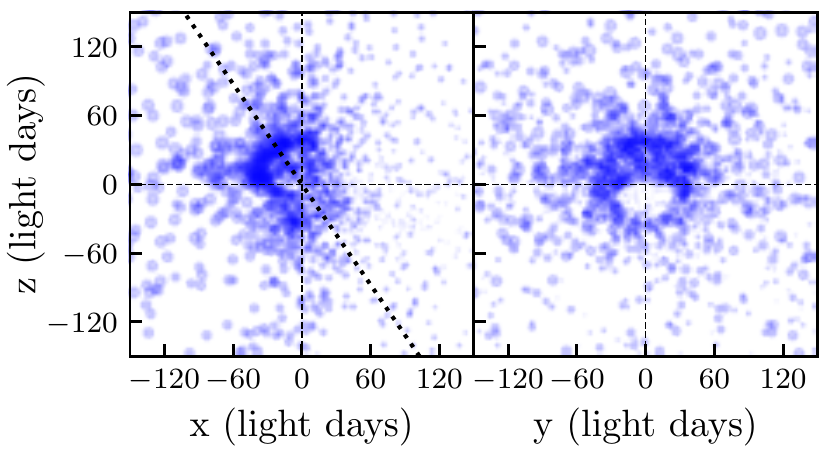}  
\caption{
One possible geometry for the \civ\ BLR, corresponding to the model shown in Figure \ref{fig:display}. Each circle represents one of the BLR test particles, and the size of the circle corresponds to the relative flux contribution of the particle. The angled dotted line shows the midplane of the BLR. The observer is situated on the positive $x$-axis.
\label{fig:geometry}}
\end{center}
\end{figure}

\subsection{BLR Model Parameters}
\label{sect:params}

Examining the posterior probability density functions (PDFs) for the model parameters, we find two clusters of solutions: those with $\theta_i$ and $\theta_o$ around 45 degrees and those with $\theta_i$ and $\theta_o$ around 85 degrees.
Due to known degeneracies resulting from the flexible parametrization of the model \citep[see, e.g.,][]{Grier++17}, different combinations of parameters can produce identical particle distributions and velocities.
For $\theta_i \sim \theta_o \rightarrow 90$ deg, the particles are arranged in a sphere, but as $\gamma \rightarrow 5$, they are increasingly concentrated along the axis of the sphere which is close to perpendicular to the observer's line-of-sight.
Since RM data cannot resolve rotations in the plane of the sky, these solutions are equivalent to near-face-on thick disk models and are thus degenerate with the first family of solutions.
The 2D posterior PDFs reveal that this is the case for the models in our sample with $\theta_i,\theta_o\sim 85$.
Unfortunately, the other model parameters are difficult to interpret in this arrangement, so we choose to exclude solutions with $\theta_i > 65$ deg.

Taking the median and 68\% confidence intervals for each parameter, we find a thick disk BLR with $\theta_o = \parthetao$ deg that is inclined by $\theta_i = \parthetai$ deg.
The parameter determining if emission is concentrated on the faces of the disk is not constrained for these models ($\gamma = \pargamma$).
The radial distribution of BLR emission drops off with radius faster than exponentially ($\beta = \parbeta$), is shifted from the origin by a minimum radius $r_{\rm min} = \parrmin$ ld, and has a median radius $r_{\rm median} = \parrmedian$ ld.
This is visible in Figure \ref{fig:geometry} with the shell-like concentration of points around the BLR center and a rapid drop-off in point density at larger radii.
The particles preferentially emit back towards the ionizing source ($\kappa \parkappa$), which agrees with photoionization predictions, and the disk midplane is found to be mostly transparent ($\xi \parxi$).

\begin{figure}[h!]
\begin{center}
\includegraphics[width=3.3in]{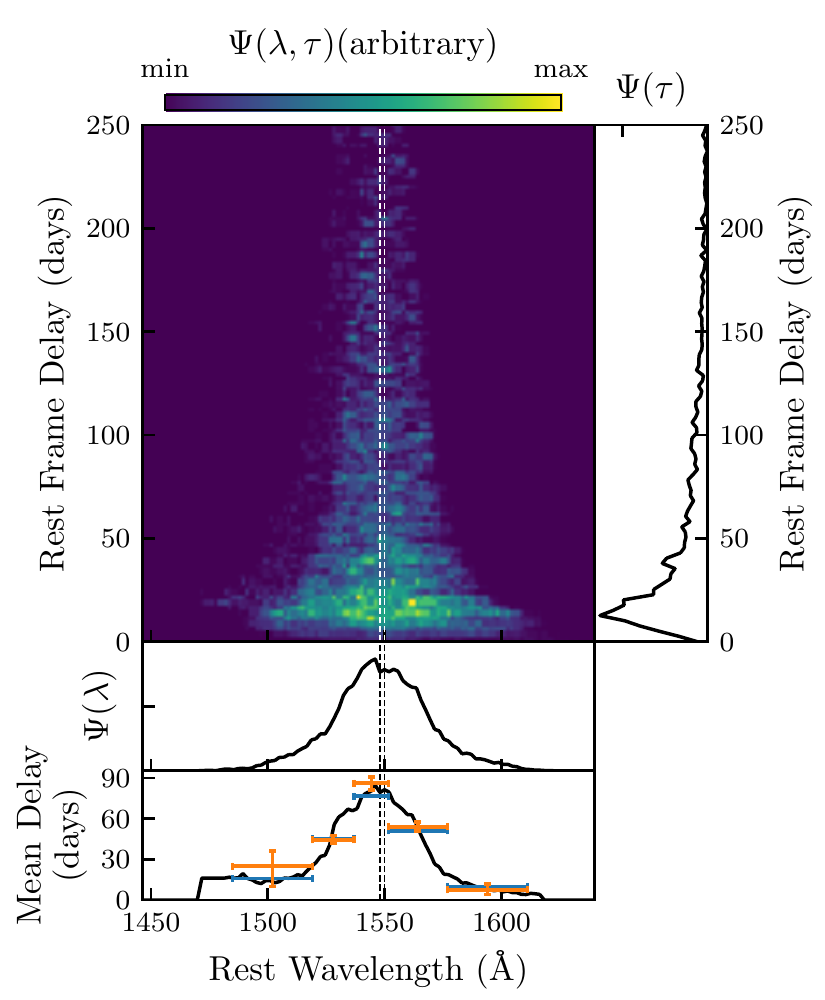}  
\caption{
Transfer function, $\Psi(\lambda,\tau)$, illustrating how continuum ($C$) variations lead to variations in the emission line ($L$): $L(\lambda, t) = \int \Psi(\lambda, \tau) C(t - \tau)d\tau$.
The bottom panels show the lag-integrated transfer function, $\Psi(\lambda)$, and the mean delay in each wavelength bin.
In orange, we show the velocity-resolved lag measurements from \citetalias{Williams++20b}, where the $x$-axis error bars denote the wavelength window.
In blue, we show the mean delay based on the model for the same wavelength windows.
Note that in the second bin, the blue model point is difficult to see since it is mostly hidden behind the orange data point.
\label{fig:transfer}}
\end{center}
\end{figure}

Kinematically, models are preferred in which most particles ($f_{\rm ellip} = \parfellip$) are on near-circular orbits, which is reflected in the symmetric nature of the transfer function (Figure \ref{fig:transfer}).
The remaining particles have infalling trajectories ($f_{\rm flow} = \parfflow$, $\theta_e = \parthetae$).
Finally, the black hole mass is very well determined at $\log_{10}(M_{\rm BH}/M_{\odot}) = \parlogmbh$.
For comparison, \citetalias{Williams++20b} measured $\log_{10}(M_{\rm BH}/M_{\odot}) = 8.63\pm 0.27$ using the lag from cross-correlation paired with a scale factor $f$ (Equation \ref{eq:revmap}) converted from \Hb-based measurements.
The modeling approach is more reliable since it does not depend on \Hb-to-\civ\ conversions for $f$, and is more precise since it avoids the additional uncertainty introduced by the intrinsic scatter in $f$.

The rest-frame time lag one would measure based on these models is $\tau_{\rm median} = \partaumedian$ days, which agrees very well with the cross-correlation measurement of $\tau_{\rm cen} = 36.5^{+2.9}_{-3.9}$ by \citetalias{Williams++20b}.
For the BLR model shown in Figure \ref{fig:transfer}, we also compute the mean emission-line lag in the five wavelength windows used to measure the velocity-resolved lags in \citetalias{Williams++20b}.
We should note that the lag for the central wavelength bin in \citetalias{Williams++20b} was based on a narrow-band filter transmission curve, rather than the top-hat function used here.
Since the transmission is higher near the center of the bin where the lags are longest, this will bias the filter-based measurements towards longer lags.
Regardless, all of our model-based measurements are consistent with the velocity-resolved lags based on cross correlation. 

\begin{figure}[h!]
\begin{center}
\includegraphics[width=3.3in]{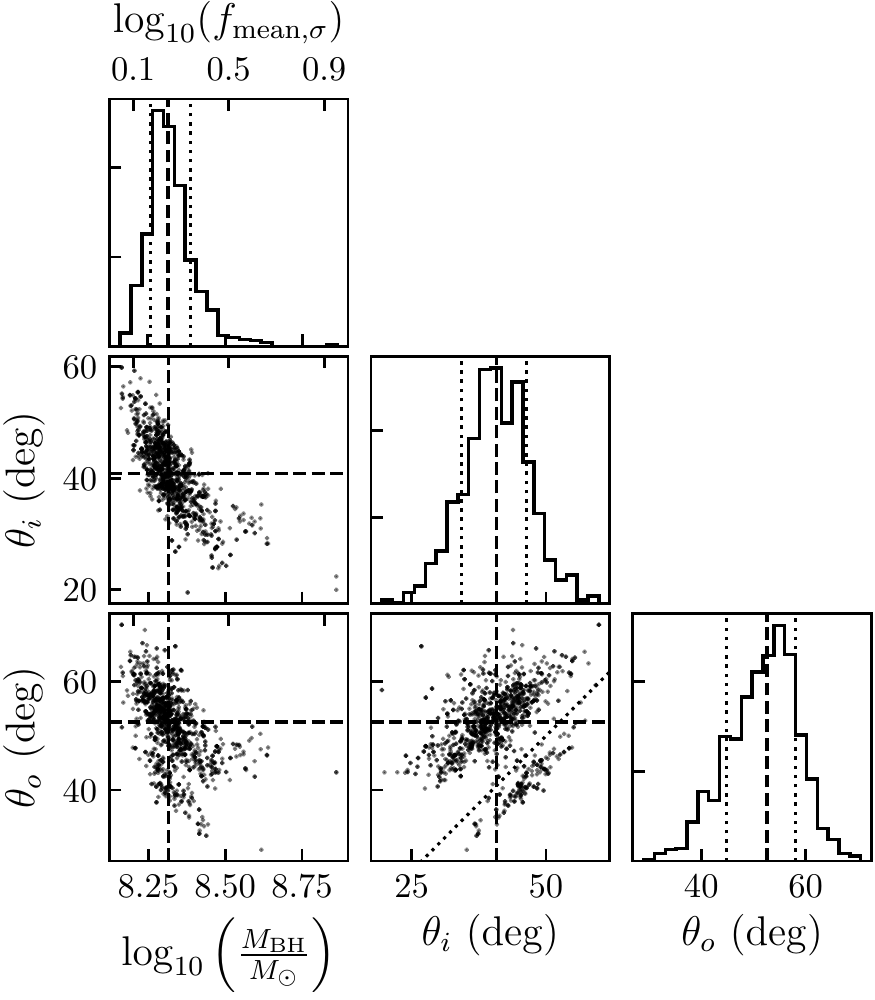}  
\caption{
2D posterior distributions for the black hole mass, inclination angle, and opening angle.
The median and 68\% confidence intervals are indicated by the dashed and dotted lines, respectively.
The corresponding scale factor, $f$, is shown on the top axes of the \mbh\ panels, assuming a line dispersion $\sigma = 4261~{\rm km~s}^{-1}$ and lag $\tau_{\rm cen} = 36.5$ days.
The angled dotted line in the $\theta_o$ vs. $\theta_i$ panel shows $\theta_o = \theta_i$.
Note that the uncertainties on $\sigma$ and $\tau_{\rm cen}$ are not included in this conversion from $\log_{10}(M_{\rm BH}/M_\odot)$ to $\log_{10} f$, so the true posterior distribution for $\log_{10} f$ is slightly broader.
\label{fig:2d_posterior}}
\end{center}
\end{figure}

In Figure \ref{fig:2d_posterior}, we show the 2D posteriors for $\log_{10}(M_{\rm BH}/M_{\odot})$, $\theta_i$, and $\theta_o$.
There is an anti-correlation between the black hole mass and inclination angle, which is expected---as the BLR is tilted closer to face on, a larger black hole is required to reproduce the observed line width. 
We also find two streaks of solutions that can be separated by the line $\theta_i = \theta_o$.
Those with $\theta_i > \theta_o$ tend to have slightly more transparent midplanes and a higher fraction of particles on near-circular orbits, but all other parameter distributions remain indistinguishable.
Importantly, both cases give the same black hole mass.

\subsection{Scale factor $f$}

Using the \mbh\ posterior PDF from our BLR model along with the \civ\ line widths and lags measured in \citetalias{Williams++20b}, we can compute the appropriate scale factor, $f$, for the SDSS J2222+2745 BLR.
We use the rest-frame lag, $\tau_{\rm cen} = 36.5^{+2.9}_{-3.9}$ days, and the four emission-line widths: $\Delta V_{{\rm mean},{\rm FWHM}} = 7734 \pm 59~{\rm km~s}^{-1}$, $\Delta V_{{\rm mean},\sigma} = 4261\pm 49~{\rm km~s}^{-1}$, $\Delta V_{{\rm rms},{\rm FWHM}} = 9219\pm 458~{\rm km~s}^{-1}$, and $\Delta V_{{\rm rms},\sigma} = 5907\pm 148~{\rm km~s}^{-1}$.

We propagate all uncertainties utilizing the full \mbh\ posterior as follows:
For each sample in the posterior, we draw an emission-line width from a normal distribution, $\mathcal{N}(\Delta V, \sigma_{\Delta V}^2)$, where $\Delta V$ is the median value reported above and $\sigma_{\Delta V}$ is the corresponding uncertainty.
We then draw a time lag using the same approach, assuming an uncertainty that is equal to the average of the upper and lower uncertainties, $3.4$ days.
From this, we construct a sample of $\log_{10} f$ values and compute the median and 68\% confidence intervals.
We find $\log_{10}(f_{{\rm mean},{\sigma}}) = \fmeansigma$, $\log_{10}(f_{{\rm mean, FWHM}}) = \fmeanfwhm$, $\log_{10}(f_{{\rm rms},{\sigma}}) = \frmssigma$, and $\log_{10}(f_{{\rm rms, FWHM}}) = \frmsfwhm$.
The uncertainties on $\log_{10}f$ are larger than those for $\log_{10}(M_{\rm BH}/M_\odot)$ since they include the uncertainties on $\Delta V$ and $\tau$.

\begin{figure}[h!]
\begin{center}
\includegraphics[width=3.3in]{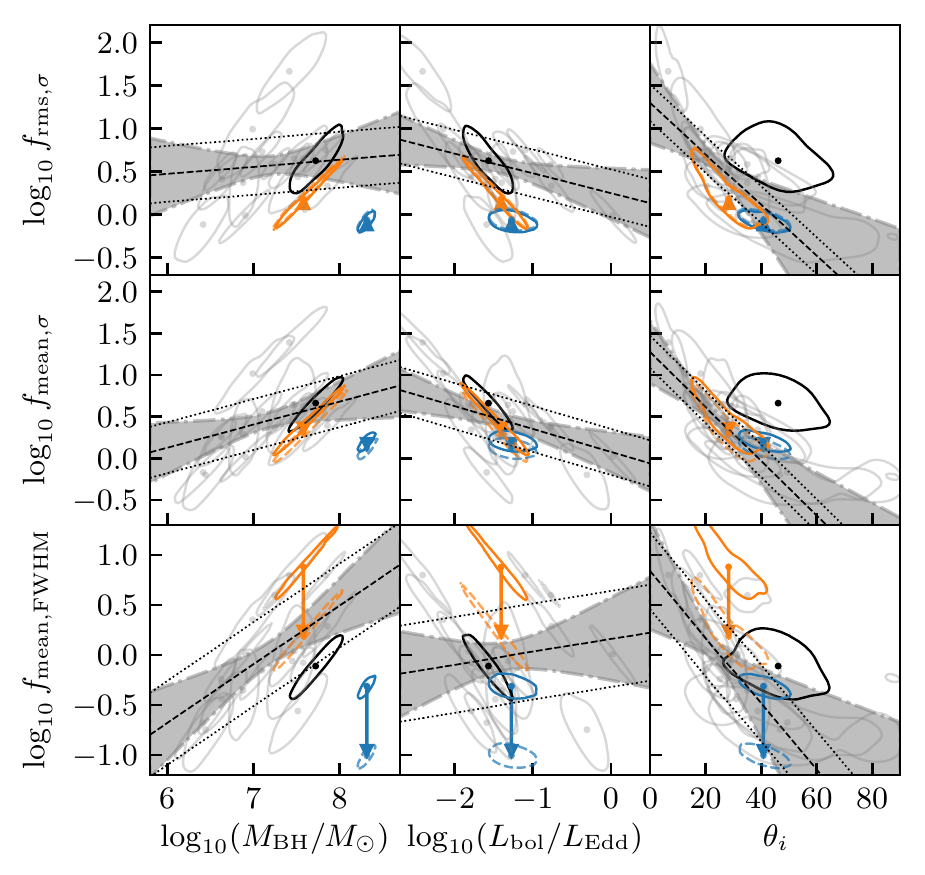}  
\caption{
Correlations between the scale factor $f$ and BLR model parameters and AGN properties.
Each point and contour pair shows the median and 68\% confidence region of the 2D posterior PDFs.
The blue point is SDSS J2222+2745, orange (black) is the \civ\ (\Hb) BLR of NGC 5548 \citep{Williams++20a}, and the grey points and fits are those for the \Hb\ BLR from \citet{Williams++18}.
For the two \civ\ measurements, we also show a shifted contour (dashed) based on the $f_{{\rm H}\beta}$ to $f_{\rm CIV}$ conversions calculated in \citetalias{Williams++20b}.
\label{fig:potato}}
\end{center}
\end{figure}

In Figure \ref{fig:potato}, we show the scale factor plotted against various BLR model and AGN parameters, along with the data and fits for the \Hb\ BLR by \citet{Williams++18} and the \civ\ and \Hb\ BLRs for NGC 5548 \citep{Williams++20a}.
To determine the bolometric luminosity, we use $\log_{10}(L_{1350}/{\rm erg~s}^{-1}) = 44.66 \pm 0.18$ from \citetalias{Williams++20b} and assume a bolometric correction of $BC_{1350} = 3.81$, computed by \citet{Shen++11} using the composite quasar spectral energy distributions of \citet{Richards++06}.

\citetalias{Williams++20b} also computed conversion factors between $f_{{\rm H}\beta}$ and $f_{{\rm CIV}}$ by comparing AGNs with line width and lag measurements available for both \Hb\ and \civ.
They found that for a relation of the form $\log_{10}(f_{\rm CIV}) = \log_{10} (f_{{\rm H}\beta}) + \alpha$, the best fits are $\alpha_{{\rm mean},\sigma} = 0.087 \pm 0.007$, $\alpha_{{\rm rms},\sigma} = -0.021 \pm 0.028$, and $\alpha_{{\rm mean},{\rm FWHM}} = 0.694 \pm 0.008$.
In Figure \ref{fig:potato}, we also show the two \civ-based contours shifted by these conversion factors.

An anti-correlation between $f$ and $\theta_i$ is expected to exist, regardless of the emission line or AGN properties.
The parameter $\theta_i$ simply describes the orientation of the BLR in the sky, so for a BLR model with disk-like kinematics, as $\theta_i$ increases, $f$ must decrease to account for the larger line-of-sight velocity component contributing to the emission-line width.
The SDSS J2222+2745 measurement happens to fall on the same relation as the \Hb\ BLR measurements, but it is possible that \civ\ BLRs follow a different relation and this is a chance alignment.
A larger sample of \civ\ BLR models is necessary to distinguish between the two options.

While $\theta_i$ itself is not an observable, it may be deduced in AGNs with radio jets if the BLR rotation axis and jet are aligned.
BLR modeling of 3C 120 \citep{Grier++17} and the spatially resolved BLR of 3C 273 from interferometry \citep{Gravity++18} support that this is the case, although a larger sample size is still necessary to determine if this relation holds for every AGN.

We also find that the SDSS J2222+2745 measurement lies below the \Hb\ relations with \mbh, and the offset is exaggerated by the $f_{\rm CIV}$ to $f_{{\rm H}\beta}$ correction factor.
The \Hb\ relation between $f_{{\rm mean}, {\rm FWHM}}$ and \mbh\ is currently detected only at the 2$\sigma$ level \citep{Williams++18}, so a larger sample will be necessary to solidify the relationship (Villafana et al. 2021, in prep).
Assuming that the relationship is real for the \Hb\ BLR, the offset of SDSS J2222+2745 could mean that the correction factors do not hold for all AGNs, or that there is something inherently different about the SDSS J2222+2745 BLR.
One possibility is that \civ\ BLRs follow a different relationship than \Hb\ BLRs, although the NGC 5548 measurement is better aligned with the \Hb-based measurements.
Another possibility is that BLR environments are different at $z=2.805$ than at $z<0.1$.
Determining the source of the offset will require additional modeling of BLRs using multiple emission lines and across a wider range of redshifts.

\section{Conclusions}
\label{sect:conclusions}

The results presented here are the first of their kind for an AGN at $z>0.1$ and at the peak of AGN activity.
The tight constraints on \mbh\ and other BLR properties demonstrate that gravitational lensing is a powerful tool that can provide data good enough for detailed BLR studies.
While opportunities like that of SDSS J2222+2745 are rare, they serve an important role in expanding our understanding of the BLR outside of the local Universe.

The main results of our analysis can be summarized as follows:
\begin{enumerate}
\item The \civ-emitting BLR for SDSS J2222+2745 is a thick disk with a size $r_{\rm median} = \parrmedian$ light days.
The kinematics are dominated by near-circular Keplerian motion, with the remaining 10-20\% of emission indicating infalling trajectories.
\item The median rest-frame lag produced by the BLR models is $\partaumedian$ days which agrees closely with the $\tau_{\rm cen} = 36.5^{+2.9}_{-1.9}$ day measurement from cross correlation.
The velocity-binned mean lags from the model are also consistent with the velocity-resolved lags from cross correlation.
\item The black hole mass for SDSS J2222+2745 is $\log_{10}(M_{\rm BH}/M_\odot) = \parlogmbh$.
This corresponds to scale factors of $\log_{10}(f_{{\rm mean},{\sigma}}) = \fmeansigma$, $\log_{10}(f_{{\rm mean, FWHM}}) = \fmeanfwhm$, $\log_{10}(f_{{\rm rms},{\sigma}}) = \frmssigma$, and $\log_{10}(f_{{\rm rms, FWHM}}) = \frmsfwhm$.
\end{enumerate}

\acknowledgments

The data presented here are based in part on observations obtained at the international Gemini Observatory, a program of NSF’s NOIRLab, which is managed by the Association of Universities for Research in Astronomy (AURA) under a cooperative agreement with the National Science Foundation on behalf of the Gemini Observatory partnership: the National Science Foundation (United States), National Research Council (Canada), Agencia Nacional de Investigaci\'{o}n y Desarrollo (Chile), Ministerio de Ciencia, Tecnolog\'{i}a e Innovaci\'{o}n (Argentina), Minist\'{e}rio da Ci\^{e}ncia, Tecnologia, Inova\c{c}\~{o}es e Comunica\c{c}\~{o}es (Brazil), and Korea Astronomy and Space Science Institute (Republic of Korea).
The Gemini data were obtained from programs GN-2016B-Q-28, GN-2017A-FT-9, GN-2017B-Q-33, GN-2018A-Q-103, GN-2018B-Q-143, GN-2019A-Q-203, GN-2019B-Q-232, GN-2020A-Q-105, and GN-2020B-Q-132 (PI Treu), and were processed using the Gemini IRAF package.

The data presented here were obtained in part with ALFOSC, which is provided by the Instituto de Astrofisica de Andalucia (IAA) under a joint agreement with the University of Copenhagen and NOTSA.
Partly based on observations made with the Nordic Optical Telescope, operated by the Nordic Optical Telescope Scientific Association at the Observatorio del Roque de los Muchachos, La Palma, Spain, of the Instituto de Astrofisica de Canarias.

This work was enabled by observations made from the Gemini North telescope, located within the Maunakea Science Reserve and adjacent to the summit of Maunakea. We are grateful for the privilege of observing the Universe from a place that is unique in both its astronomical quality and its cultural significance.

This research made use of Astropy,\footnote{http://www.astropy.org} a community-developed core Python package for Astronomy \citep{astropy:2013, astropy:2018}

PW and TT gratefully acknowledge support by the National Science Foundation through grant AST-1907208 ``Collaborative Research: Establishing the foundations of black hole mass measurements of AGN across cosmic time''  and by the Packard Foundation through a Packard Research Fellowship to TT.
Research at UC Irvine was supported by NSF grant AST-1907290. 
KH acknowledges support from STFC grant ST/R000824/1.

\software{Astropy \citep{astropy:2013, astropy:2018}, {\sc DNest4} \citep{dnest4}, IRAF \citep{Tody86,Tody93}, Scipy \citep{scipy}}


\begin{thebibliography}{}
\expandafter\ifx\csname natexlab\endcsname\relax\def\natexlab#1{#1}\fi

\bibitem[{{Astropy Collaboration} {et~al.}(2013){Astropy Collaboration},
  {Robitaille}, {Tollerud}, {Greenfield}, {Droettboom}, {Bray}, {Aldcroft},
  {Davis}, {Ginsburg}, {Price-Whelan}, {Kerzendorf}, {Conley}, {Crighton},
  {Barbary}, {Muna}, {Ferguson}, {Grollier}, {Parikh}, {Nair}, {Unther},
  {Deil}, {Woillez}, {Conseil}, {Kramer}, {Turner}, {Singer}, {Fox}, {Weaver},
  {Zabalza}, {Edwards}, {Azalee Bostroem}, {Burke}, {Casey}, {Crawford},
  {Dencheva}, {Ely}, {Jenness}, {Labrie}, {Lim}, {Pierfederici}, {Pontzen},
  {Ptak}, {Refsdal}, {Servillat}, \& {Streicher}}]{astropy:2013}
{Astropy Collaboration}, {Robitaille}, T.~P., {Tollerud}, E.~J., {et~al.} 2013,
  \aap, 558, A33

\bibitem[{{Astropy Collaboration} {et~al.}(2018){Astropy Collaboration},
  {Price-Whelan}, {Sip{H{o}}cz}, {G{"u}nther}, {Lim}, {Crawford}, {Conseil},
  {Shupe}, {Craig}, {Dencheva}, {Ginsburg}, {Vand erPlas}, {Bradley},
  {P{'e}rez-Su{'a}rez}, {de Val-Borro}, {Aldcroft}, {Cruz}, {Robitaille},
  {Tollerud}, {Ardelean}, {Babej}, {Bach}, {Bachetti}, {Bakanov}, {Bamford},
  {Barentsen}, {Barmby}, {Baumbach}, {Berry}, {Biscani}, {Boquien}, {Bostroem},
  {Bouma}, {Brammer}, {Bray}, {Breytenbach}, {Buddelmeijer}, {Burke},
  {Calderone}, {Cano Rodr{'i}guez}, {Cara}, {Cardoso}, {Cheedella}, {Copin},
  {Corrales}, {Crichton}, {D'Avella}, {Deil}, {Depagne}, {Dietrich}, {Donath},
  {Droettboom}, {Earl}, {Erben}, {Fabbro}, {Ferreira}, {Finethy}, {Fox},
  {Garrison}, {Gibbons}, {Goldstein}, {Gommers}, {Greco}, {Greenfield},
  {Groener}, {Grollier}, {Hagen}, {Hirst}, {Homeier}, {Horton}, {Hosseinzadeh},
  {Hu}, {Hunkeler}, {Ivezi{'c}}, {Jain}, {Jenness}, {Kanarek}, {Kendrew},
  {Kern}, {Kerzendorf}, {Khvalko}, {King}, {Kirkby}, {Kulkarni}, {Kumar},
  {Lee}, {Lenz}, {Littlefair}, {Ma}, {Macleod}, {Mastropietro}, {McCully},
  {Montagnac}, {Morris}, {Mueller}, {Mumford}, {Muna}, {Murphy}, {Nelson},
  {Nguyen}, {Ninan}, {N{"o}the}, {Ogaz}, {Oh}, {Parejko}, {Parley}, {Pascual},
  {Patil}, {Patil}, {Plunkett}, {Prochaska}, {Rastogi}, {Reddy Janga},
  {Sabater}, {Sakurikar}, {Seifert}, {Sherbert}, {Sherwood-Taylor}, {Shih},
  {Sick}, {Silbiger}, {Singanamalla}, {Singer}, {Sladen}, {Sooley},
  {Sornarajah}, {Streicher}, {Teuben}, {Thomas}, {Tremblay}, {Turner},
  {Terr{'o}n}, {van Kerkwijk}, {de la Vega}, {Watkins}, {Weaver}, {Whitmore},
  {Woillez}, {Zabalza}, \& {Astropy Contributors}}]{astropy:2018}
{Astropy Collaboration}, {Price-Whelan}, A.~M., {Sip{H{o}}cz}, B.~M., {et~al.}
  2018, aj, 156, 123

\bibitem[{{Blandford} \& {McKee}(1982)}]{blandford82}
{Blandford}, R.~D., \& {McKee}, C.~F. 1982, \apj, 255, 419

\bibitem[{Brewer \& Foreman-Mackey(2016)}]{dnest4}
Brewer, B., \& Foreman-Mackey, D. 2016, Journal of Statistical Software, 86,
  doi:10.18637/jss.v086.i07

\bibitem[{{Brewer} {et~al.}(2011){Brewer}, {Treu}, {Pancoast}, {Barth},
  {Bennert}, {Bentz}, {Filippenko}, {Greene}, {Malkan}, \& {Woo}}]{brewer11b}
{Brewer}, B.~J., {Treu}, T., {Pancoast}, A., {et~al.} 2011, \apjl, 733, L33

\bibitem[{{Dahle} {et~al.}(2013){Dahle}, {Gladders}, {Sharon}, {Bayliss},
  {Wuyts}, {Abramson}, {Koester}, {Groeneboom}, {Brinckmann}, {Kristensen},
  {Lindholmer}, {Nielsen}, {Krogager}, \& {Fynbo}}]{Dahle++13}
{Dahle}, H., {Gladders}, M.~D., {Sharon}, K., {et~al.} 2013, \apj, 773, 146

\bibitem[{{Ding} {et~al.}(2020){Ding}, {Silverman}, {Treu}, {Schulze},
  {Schramm}, {Birrer}, {Park}, {Jahnke}, {Bennert}, {Kartaltepe}, {Koekemoer},
  {Malkan}, \& {Sanders}}]{Ding++20}
{Ding}, X., {Silverman}, J., {Treu}, T., {et~al.} 2020, \apj, 888, 37

\bibitem[{{Dyrland}(2019)}]{Dyrland19}
{Dyrland}, K. 2019, Master's thesis, University of Oslo,
  http://urn.nb.no/URN:NBN:no-73119

\bibitem[{{Ferrarese} \& {Ford}(2005)}]{ferrarese05}
{Ferrarese}, L., \& {Ford}, H. 2005, \ssr, 116, 523

\bibitem[{{Ferrarese} \& {Merritt}(2000)}]{ferrarese00}
{Ferrarese}, L., \& {Merritt}, D. 2000, \apjl, 539, L9

\bibitem[{{Gebhardt} {et~al.}(2000){Gebhardt}, {Bender}, {Bower}, {Dressler},
  {Faber}, {Filippenko}, {Green}, {Grillmair}, {Ho}, {Kormendy}, {Lauer},
  {Magorrian}, {Pinkney}, {Richstone}, \& {Tremaine}}]{gebhardt00}
{Gebhardt}, K., {Bender}, R., {Bower}, G., {et~al.} 2000, \apjl, 539, L13

\bibitem[{{Gravity Collaboration} {et~al.}(2018){Gravity Collaboration},
  {Sturm}, {Dexter}, {Pfuhl}, {Stock}, {Davies}, {Lutz}, {Cl{\'e}net},
  {Eckart}, {Eisenhauer}, {Genzel}, {Gratadour}, {H{\"o}nig}, {Kishimoto},
  {Lacour}, {Millour}, {Netzer}, {Perrin}, {Peterson}, {Petrucci}, {Rouan},
  {Waisberg}, {Woillez}, {Amorim}, {Brandner}, {F{\"o}rster Schreiber},
  {Garcia}, {Gillessen}, {Ott}, {Paumard}, {Perraut}, {Scheithauer},
  {Straubmeier}, {Tacconi}, \& {Widmann}}]{Gravity++18}
{Gravity Collaboration}, {Sturm}, E., {Dexter}, J., {et~al.} 2018, \nat, 563,
  657

\bibitem[{{Grier} {et~al.}(2017){Grier}, {Pancoast}, {Barth}, {Fausnaugh},
  {Brewer}, {Treu}, \& {Peterson}}]{Grier++17}
{Grier}, C.~J., {Pancoast}, A., {Barth}, A.~J., {et~al.} 2017, \apj, 849, 146

\bibitem[{{Hook} {et~al.}(2004){Hook}, {J{\o}rgensen}, {Allington-Smith},
  {Davies}, {Metcalfe}, {Murowinski}, \& {Crampton}}]{GMOS}
{Hook}, I.~M., {J{\o}rgensen}, I., {Allington-Smith}, J.~R., {et~al.} 2004,
  \pasp, 116, 425

\bibitem[{{Pancoast} {et~al.}(2011){Pancoast}, {Brewer}, \&
  {Treu}}]{pancoast11}
{Pancoast}, A., {Brewer}, B.~J., \& {Treu}, T. 2011, \apj, 730, 139

\bibitem[{{Pancoast} {et~al.}(2014){Pancoast}, {Brewer}, {Treu}, {Park},
  {Barth}, {Bentz}, \& {Woo}}]{pancoast14b}
{Pancoast}, A., {Brewer}, B.~J., {Treu}, T., {et~al.} 2014, \mnras, 445, 3073

\bibitem[{{Pancoast} {et~al.}(2012){Pancoast}, {Brewer}, {Treu}, {Barth},
  {Bennert}, {Canalizo}, {Filippenko}, {Gates}, {Greene}, {Li}, {Malkan},
  {Sand}, {Stern}, {Woo}, {Assef}, {Bae}, {Buehler}, {Cenko}, {Clubb},
  {Cooper}, {Diamond-Stanic}, {Hiner}, {H{\"o}nig}, {Joner}, {Kandrashoff},
  {Laney}, {Lazarova}, {Nierenberg}, {Park}, {Silverman}, {Son}, {Sonnenfeld},
  {Thorman}, {Tollerud}, {Walsh}, \& {Walters}}]{pancoast12}
---. 2012, \apj, 754, 49

\bibitem[{{Park} {et~al.}(2012){Park}, {Kelly}, {Woo}, \& {Treu}}]{park12b}
{Park}, D., {Kelly}, B.~C., {Woo}, J.-H., \& {Treu}, T. 2012, \apjs, 203, 6

\bibitem[{{Peterson}(1993)}]{peterson93}
{Peterson}, B.~M. 1993, \pasp, 105, 247

\bibitem[{{Peterson}(2014)}]{peterson14}
---. 2014, \ssr, 183, 253

\bibitem[{{Richards} {et~al.}(2006){Richards}, {Lacy}, {Storrie-Lombardi},
  {Hall}, {Gallagher}, {Hines}, {Fan}, {Papovich}, {Vanden Berk}, {Trammell},
  {Schneider}, {Vestergaard}, {York}, {Jester}, {Anderson}, {Budav{\'a}ri}, \&
  {Szalay}}]{Richards++06}
{Richards}, G.~T., {Lacy}, M., {Storrie-Lombardi}, L.~J., {et~al.} 2006, \apjs,
  166, 470

\bibitem[{{Shen} {et~al.}(2011){Shen}, {Richards}, {Strauss}, {Hall},
  {Schneider}, {Snedden}, {Bizyaev}, {Brewington}, {Malanushenko},
  {Malanushenko}, {Oravetz}, {Pan}, \& {Simmons}}]{Shen++11}
{Shen}, Y., {Richards}, G.~T., {Strauss}, M.~A., {et~al.} 2011, \apjs, 194, 45

\bibitem[{{Tody}(1986)}]{Tody86}
{Tody}, D. 1986, in Society of Photo-Optical Instrumentation Engineers (SPIE)
  Conference Series, Vol. 627, Instrumentation in astronomy VI, ed. D.~L.
  {Crawford}, 733

\bibitem[{{Tody}(1993)}]{Tody93}
{Tody}, D. 1993, in Astronomical Society of the Pacific Conference Series,
  Vol.~52, Astronomical Data Analysis Software and Systems II, ed. R.~J.
  {Hanisch}, R.~J.~V. {Brissenden}, \& J.~{Barnes}, 173

\bibitem[{Virtanen {et~al.}(2020)Virtanen, Gommers, Oliphant, Haberland, Reddy,
  Cournapeau, Burovski, Peterson, Weckesser, Bright, {van der Walt}, Brett,
  Wilson, Millman, Mayorov, Nelson, Jones, Kern, Larson, Carey, Polat, Feng,
  Moore, {VanderPlas}, Laxalde, Perktold, Cimrman, Henriksen, Quintero, Harris,
  Archibald, Ribeiro, Pedregosa, {van Mulbregt}, \& {SciPy 1.0
  Contributors}}]{scipy}
Virtanen, P., Gommers, R., Oliphant, T.~E., {et~al.} 2020, Nature Methods, 17,
  261

\bibitem[{{Williams} {et~al.}(2018){Williams}, {Pancoast}, {Treu}, {Brewer},
  {Barth}, {Bennert}, {Buehler}, {Canalizo}, {Cenko}, {Clubb}, {Cooper},
  {Filippenko}, {Gates}, {Hoenig}, {Joner}, {Kandrashoff}, {Laney}, {Lazarova},
  {Li}, {Malkan}, {Rex}, {Silverman}, {Tollerud}, {Walsh}, \&
  {Woo}}]{Williams++18}
{Williams}, P.~R., {Pancoast}, A., {Treu}, T., {et~al.} 2018, \apj, 866, 75

\bibitem[{{Williams} {et~al.}(2020{\natexlab{a}}){Williams}, {Pancoast},
  {Treu}, {Brewer}, {Peterson}, {Barth}, {Malkan}, {De Rosa}, {Horne}, {Kriss},
  {Arav}, {Bentz}, {Cackett}, {Dalla Bont{\`a}}, {Dehghanian}, {Done},
  {Ferland}, {Grier}, {Kaastra}, {Kara}, {Kochanek}, {Mathur}, {Mehdipour},
  {Pogge}, {Proga}, {Vestergaard}, {Waters}, {Adams}, {Anderson},
  {Ar{\'e}valo}, {Beatty}, {Bennert}, {Bigley}, {Bisogni}, {Borman}, {Boroson},
  {Bottorff}, {Brandt}, {Breeveld}, {Brotherton}, {Brown}, {Brown}, {Canalizo},
  {Carini}, {Clubb}, {Comerford}, {Corsini}, {Crenshaw}, {Croft}, {Croxall},
  {Deason}, {De Lorenzo-C{\'a}ceres}, {Denney}, {Dietrich}, {Edelson},
  {Efimova}, {Ely}, {Evans}, {Fausnaugh}, {Filippenko}, {Flatland}, {Fox},
  {Gardner}, {Gates}, {Gehrels}, {Geier}, {Gelbord}, {Gonzalez}, {Gorjian},
  {Greene}, {Grupe}, {Gupta}, {Hall}, {Henderson}, {Hicks}, {Holmbeck},
  {Holoien}, {Hutchison}, {Im}, {Jensen}, {Johnson}, {Joner}, {Jones}, {Kaspi},
  {Kelly}, {Kennea}, {Kim}, {Kim}, {Kim}, {King}, {Klimanov}, {Knigge},
  {Krongold}, {Lau}, {Lee}, {Leonard}, {Li}, {Lira}, {Lochhaas}, {Ma},
  {MacInnis}, {Manne-Nicholas}, {Mauerhan}, {McGurk}, {McHardy}, {Montuori},
  {Morelli}, {Mosquera}, {Mudd}, {M{\"u}ller-S{\'a}nchez}, {Nazarov}, {Norris},
  {Nousek}, {Nguyen}, {Ochner}, {Okhmat}, {Papadakis}, {Parks}, {Pei}, {Penny},
  {Pizzella}, {Poleski}, {Pott}, {Rafter}, {Rix}, {Runnoe}, {Saylor},
  {Schimoia}, {Scott}, {Sergeev}, {Shappee}, {Shivvers}, {Siegel}, {Simonian},
  {Siviero}, {Skielboe}, {Somers}, {Spencer}, {Starkey}, {Stevens}, {Sung},
  {Tayar}, {Tejos}, {Turner}, {Uttley}, {Van Saders}, {Vaughan}, {Vican},
  {Villanueva}, {Villforth}, {Weiss}, {Woo}, {Yan}, {Young}, {Yuk}, {Zheng},
  {Zhu}, \& {Zu}}]{Williams++20a}
---. 2020{\natexlab{a}}, \apj, 902, 74

\bibitem[{{Williams} {et~al.}(2020{\natexlab{b}}){Williams}, {Treu}, {Dahle},
  {Valenti}, {Abramson}, {Barth}, {Gladders}, {Horne}, \&
  {Sharon}}]{Williams++20b}
{Williams}, P.~R., {Treu}, T., {Dahle}, H., {et~al.} 2020{\natexlab{b}}, arXiv
  e-prints, arXiv:2011.02007

\end{thebibliography}

\end{document}